\documentclass[aps,reprint,groupedaddress,superscriptaddress,pra]{revtex4-2}
\usepackage[T1]{fontenc}
\usepackage[utf8]{inputenc}
\usepackage{newtxtext,newtxmath}
\usepackage{bm,mathtools,graphicx}
\usepackage{CJK}
\usepackage[colorlinks=true,citecolor=red,linkcolor=blue,urlcolor=blue]{hyperref}
\newcommand{\kb}{k_{\mathrm B}}
\newcommand{\avg}[1]{\langle #1\rangle}
\newcommand{\pd}{\partial}

\let\oldemph\emph
\renewcommand{\emph}[1]{\textcolor{blue}{\oldemph{#1}}}
\usepackage{xcolor}

\begin{document}
\begin{CJK*}{UTF8}{gbsn}
\title{Geodesics and a Hubble Relation in Nonstationary Thermodynamic Geometry}
\author{Wanyan Chen(陈婉妍)}
\affiliation{School of Physics and Astronomy, Beijing Normal University, Beijing, 100875, China}
\affiliation{Key Laboratory of Multiscale Spin Physics (Ministry of Education), Beijing Normal University, Beijing 100875, China}
\author{Yu-Han Ma(马宇翰)}
\email{yhma@bnu.edu.cn}
\affiliation{School of Physics and Astronomy, Beijing Normal University, Beijing, 100875, China}
\affiliation{Key Laboratory of Multiscale Spin Physics (Ministry of Education), Beijing Normal University, Beijing 100875, China}
\begin{abstract}

Thermodynamic geometry offers a versatile lens from equilibrium to nonequilibrium, yet typically assumes a stationary metric. Here we study nonstationary thermodynamic geometry, where the metric evolves even at fixed control parameters. The resulting temporal connection generalizes the geodesic equation and redistributes dissipation along optimal control paths. We further establish a thermodynamic Hubble relation that distinguishes relaxation toward equilibrium from geometric expansion or contraction. Brownian examples demonstrate that externally modulated friction can reduce minimum dissipation, while heat exchange with a finite reservoir can induce geometric contraction and Mpemba-type relaxation. Our framework lays a geometric foundation for understanding complex evolving nonequilibrium systems.

\end{abstract}
\maketitle
\emph{Introduction.---}Driving a thermodynamic system in finite time generally incurs excess dissipation
~\cite{Andresen1977,SchmiedlSeifert2007,Berut2012,Ma2020PRL}. Thermodynamic geometry originated with the
equilibrium metrics of Weinhold and Ruppeiner,
which describe thermodynamic response
and fluctuations
~\cite{Weinhold1975,Ruppeiner1979,Ruppeiner1995}.
Near equilibrium, thermodynamic length was
introduced to characterize finite-time
dissipation geometrically
~\cite{SalamonBerry1983,NultonSalamon1985},
with subsequent connections to Fisher
information and nonequilibrium work
fluctuations
~\cite{Crooks2007,FengCrooks2009,Ito2018PRL}.
In the slow-driving regime, minimizing
dissipation becomes a geodesic problem:
optimal protocols follow shortest geodesics
of the friction metric at constant excess
power~\cite{SivakCrooks2012,Zulkowski2012}.
Building on these foundations, thermodynamic
geometry has become a powerful framework for
finite-time control across diverse systems
~\cite{ZulkowskiQuantum2015,Scandi2019,
RotskoffCrooks2015,EglintonEtAl2023},
with applications ranging from heat-engine
optimization
~\cite{BrandnerSaito2020,FrimDeWeese2022,
Chen2022HeatEngine,MaFu2025}
and information erasure
~\cite{MaChenSunDong2022,ScandiEtAl2022}
to particle separation
~\cite{ChenZhaiSunDong2023,ZhaoTuMa2024}.
Further advances have connected thermodynamic
friction to optimal transport
~\cite{Aurell2011,ZhongDeWeese2024,OikawaEtAl2025},
network geometry~\cite{SawchukGraph2026},
and broader control manifolds
~\cite{SawchukSivak2026}.
Nevertheless, these approaches generally
assume a stationary thermodynamic geometry,
with the friction metric prescribed as a
fixed function of the control coordinates.

In realistic systems, however, a thermodynamic
description retains only a limited set of
variables from the full system and its
environment
~\cite{Esposito2012,FalascoEsposito2025RMP}.
Unresolved degrees of freedom can retain
memory and modify the effective response
of observable variables
~\cite{EspositoVdB2010,Seifert2012}. Such effects appear in fluctuating diffusivity~\cite{ChubynskySlater2014,ChechkinEtAl2017}
and memory-dependent cooling~\cite{LinTuMa2026}, while recent experiments have demonstrated work extraction from
hidden environmental memory~\cite{MurugaGinotLoosBechinger2026}. The Mpemba effect further illustrates how relaxation under identical external conditions can depend sensitively on initial preparation
~\cite{TezaEtAl2026,LuRaz2017,Klich2019,KumarBechhoefer2020,VuHayakawa2025}. Recent geometric analyses have also related
relaxation modes to thermodynamic stiffness
and dissipation~\cite{WangZhaoDengMa2026}.
These developments suggest that the effective
kinetic resistance of an observable system
may evolve with its hidden internal or
environmental state, even when the external
control parameters remain unchanged.

\begin{figure}[tbh!]
\centering
\includegraphics[width=\columnwidth]
{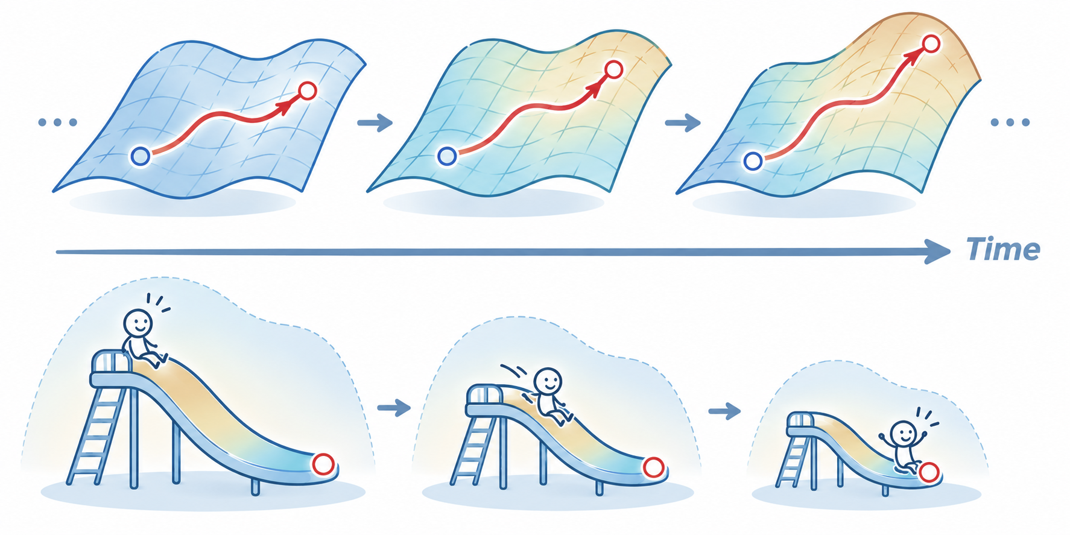}
\caption{Schematic of driving (upper)
and relaxation (lower) processes
in nonstationary thermodynamic geometry.}
\label{fig:geometry_evolution}
\end{figure}

A fundamental question then arises:
how can nonequilibrium processes be
characterized geometrically when the
accessible thermodynamic space itself
evolves in time?
As illustrated in Fig.~\ref{fig:geometry_evolution},
such evolution influences both driving
(upper) and relaxation (lower) processes. In this Letter, we develop a framework of
\textit{nonstationary thermodynamic geometry},
where evolving internal or environmental
degrees of freedom reshape the accessible
geometry. This evolution can be externally driven
or autonomous. Under time-scale separation, we construct
the evolving metric from conditional
force correlations without resolving
all hidden microscopic dynamics.
The resulting temporal connection
modifies minimum-dissipation geodesics
and redistributes excess power.
We further establish a thermodynamic
Hubble relation connecting metric evolution
to spontaneous relaxation.
Together, these results provide a
foundation for nonequilibrium thermodynamics
in evolving geometries.

\emph{Thermodynamic metrics with hidden degrees of freedom.--}Consider microscopic configurations $z$, accessible
control parameters $\bm\lambda=(\lambda^1,\ldots,\lambda^d)$,
and a hidden background variable $\alpha$.
For each fixed $(\bm\lambda,\alpha)$, let
$\pi_{\bm\lambda,\alpha}(z)$ denote the conditional
equilibrium distribution.
Two complementary metrics can be defined on the
accessible control space.
The equilibrium Fisher--Rao metric
$\mathcal I_{ij}=\langle s_i s_j\rangle_\pi$,
with $s_i=\pd_{\lambda^i}\ln\pi$, measures
statistical distinguishability
~\cite{Fisher1922,Rao1945,AmariNagaoka2000}.
Finite-time dissipation, by contrast, is described
by a friction metric obtained from the Green--Kubo
relation~\cite{Kirkwood1946,Zwanzig1964,SivakCrooks2012},
\begin{equation}
g_{ij}(\bm\lambda,\alpha)
=\beta\int_0^\infty ds\,
\avg{\delta X_i(0)\delta X_j(s)}
_{\bm\lambda,\alpha},
\label{eq:metric}
\end{equation}
where $X_i=-\pd_{\lambda^i}E$,
$\delta X_i=X_i-\langle X_i\rangle$,
and $\beta=(\kb T_b)^{-1}$.
The correlations are evaluated with both parameters
held fixed. Under detailed balance and integrable
correlations, $g_{ij}$ is symmetric and positive
semidefinite, defining a friction metric on
nondegenerate control directions.
For a canonical equilibrium family,
$s_i=\beta\delta X_i$, so that
$g_{ij}=\kb T_b\int_0^\infty ds\,
\langle s_i(0)s_j(s)\rangle_\pi$.
Thus, Fisher information captures equal-time
fluctuations, whereas friction also reflects
their relaxation kinetics
~\cite{Crooks2007,SivakCrooks2012}.
If the relevant correlations decay exponentially
with a common time $\tau_r$, then
$g_{ij}=\kb T_b\tau_r\mathcal I_{ij}$.
A background that changes only the kinetics
therefore leaves the Fisher geometry unchanged
while modifying the friction geometry; if it
also changes the equilibrium distribution,
both metrics may evolve.

When $\alpha=\alpha(t)$ evolves through external modulation or autonomous dynamics, the conditional
metrics become time dependent.
If both the controls and background vary slowly
compared with the force-correlation time $\tau_c$,
the instantaneous frozen metrics provide a local
response approximation.
The excess power associated with accessible driving
then takes the quadratic form
~\cite{SivakCrooks2012,Zulkowski2012}
\begin{equation}
P_{\mathrm{ex}}^{(\lambda)}(t)\simeq
g_{ij}(\bm\lambda,t)
\dot\lambda^i\dot\lambda^j,
\label{eq:power}
\end{equation}
where $g_{ij}(\bm\lambda,t)
=g_{ij}[\bm\lambda,\alpha(t)]$.
This approximation requires
$\tau_c|\dot\lambda^i|/\lambda_{\rm sc}^i\ll1$
and $\tau_c|\dot\alpha|/\alpha_{\rm sc}\ll1$,
where $\lambda_{\rm sc}^i$ and
$\alpha_{\rm sc}$ denote the characteristic
parameter scales over which the local
response changes appreciably. Beyond this regime, particularly near critical
slowing down, nonlocal memory can invalidate
the instantaneous metric description.
Equation~(\ref{eq:power}) captures only the
dissipation associated with accessible driving
for a given background history.
During spontaneous relaxation the controls
remain fixed, whereas external modulation may
drive the background and controls through distinct
physical channels. In either case, the full energetic cost may also include
hidden-channel dissipation and cross-channel
responses~\cite{Esposito2012}.
The conditional metric is therefore distinct
from that obtained by freely eliminating
the hidden velocity from the full
dissipation tensor, as detailed in
\emph{Appendix~A} of the End Matter.

\emph{Generalized geodesics and minimum dissipation.--}We first consider a prescribed background history
$\alpha(t)$, independent of the accessible control
protocol. For a finite-time transformation on
the evolving geometry illustrated in
Fig.~\ref{fig:geometry_evolution} (upper),
with fixed endpoints
$\bm\lambda(0)=\bm\lambda_i$ and
$\bm\lambda(\tau)=\bm\lambda_f$,
integrating the excess power in
Eq.~(\ref{eq:power}) gives the leading
excess work in the slow-driving regime,
\begin{equation}
\mathcal W_\lambda
=\int_0^\tau dt\,
g_{ij}(\bm\lambda,t)
\dot\lambda^i\dot\lambda^j,
\label{eq:action}
\end{equation}
For stationary metrics, minimizing this functional
yields thermodynamic geodesics with constant excess
power~\cite{SalamonBerry1983,Crooks2007,
SivakCrooks2012}, a principle also underlying
geometric shortcuts to isothermality
~\cite{LiChenSunDong2022}.
For a time-dependent metric, the Euler--Lagrange
equation generalizes to
\begin{equation}
\ddot\lambda^k
+\Gamma^k_{ij}[g]\dot\lambda^i\dot\lambda^j
+g^{k\ell}\partial_tg_{\ell j}\dot\lambda^j=0,
\label{eq:geodesic}
\end{equation}
where $\Gamma^k_{ij}[g]
=\frac12g^{k\ell}(\partial_i g_{\ell j}
+\partial_j g_{\ell i}-\partial_\ell g_{ij})$
is the Levi--Civita connection of the instantaneous
metric~\cite{doCarmo1992}.
This generalized geodesic equation follows
from extremizing the quadratic dissipation
functional, consistent with the variational
theory of time-dependent Riemannian metrics
~\cite{GraciaRivasTorres2026}. Unlike stationary geodesics, its solutions need not
minimize geometric length.
The additional term $2\mathsf H^k{}_j\dot\lambda^j$
defines a temporal connection
$\mathsf H^i{}_j=\frac12g^{ik}\partial_tg_{kj}$,
whose eigenvalues describe local rates of metric
expansion or contraction.
For the instantaneous excess power
$P_{\rm ex}^{(\lambda)}
=g_{ij}\dot\lambda^i\dot\lambda^j$,
Eq.~(\ref{eq:geodesic}) gives
\begin{equation}
\frac{dP_{\rm ex}^{(\lambda)}}{dt}
=-\partial_tg_{ij}\dot\lambda^i\dot\lambda^j.
\label{eq:energy}
\end{equation}
Thus, metric evolution modifies the conventional
constant-power law. Expansion along the control
direction
($\partial_tg_{ij}\dot\lambda^i\dot\lambda^j>0$)
lowers the optimal excess power, whereas contraction
($\partial_tg_{ij}\dot\lambda^i\dot\lambda^j<0$)
increases it.

An important class of nonstationary geometries arises
when the background changes relaxation timescales
without altering equilibrium statistics.
If the equilibrium force covariance
$\langle\delta X_i\delta X_j\rangle$ is independent
of $\alpha$, while the force correlations retain
their shapes and their timescales are uniformly
rescaled by $a^2(\alpha)>0$, the substitution
$s=a^2(\alpha)u$ in the Green--Kubo
relation~(\ref{eq:metric}) gives
$g_{ij}(\bm\lambda,\alpha)
=a^2(\alpha)h_{ij}(\bm\lambda)$,
where $h_{ij}$ is the reference friction metric.
This scaling also applies to multiple relaxation
modes with unchanged relative amplitudes.
The evolving background thus generates a conformally
separable metric
$g_{ij}(\bm\lambda,t)=a^2(t)h_{ij}(\bm\lambda)$,
with $a(t)=a[\alpha(t)]$.
Its temporal connection reduces to
$\mathsf H^i{}_j=H\delta^i_j$,
where $H=\dot a/a$ is the common expansion or
contraction rate of all control directions.
Equation~(\ref{eq:geodesic}) then becomes
\begin{equation}
\ddot\lambda^k+
\Gamma^k_{ij}[h]\dot\lambda^i\dot\lambda^j
+2H\dot\lambda^k=0.
\label{eq:geodesicH}
\end{equation}
The term $2H\dot\lambda^k$ changes the speed along
the optimal path without changing its spatial shape:
the path remains a geodesic of $h_{ij}$ up to
reparameterization. Expansion ($H>0$) slows the
optimal motion along this path, whereas contraction
($H<0$) accelerates it.
Correspondingly, Eq.~(\ref{eq:energy}) gives
$\dot P_{\rm ex}^{(\lambda)}=-2HP_{\rm ex}^{(\lambda)}$,
implying $a^2(t)P_{\rm ex}^{(\lambda)}(t)
=\mathrm{const}$. Thus, the same expansion rate
controls the time parameterization and the evolution
of optimal excess power.

For this separable metric, define
$ds_h^2=h_{ij}d\lambda^i d\lambda^j$ and
$I_\tau=\int_0^\tau a^{-2}(t)\,dt$.
Applying the Cauchy--Schwarz inequality to
Eq.~(\ref{eq:action}), as in conventional
thermodynamic-length optimization
~\cite{SalamonBerry1983,NultonSalamon1985,
SivakCrooks2012}, yields
$\mathcal W_\lambda\geq
\mathcal L_h^2/I_\tau\geq\ell_h^2/I_\tau$.
Equality in the first inequality determines the
optimal protocol along a given path $\gamma$ of
length $\mathcal L_h=\int_\gamma ds_h$, while
equality in the second selects a shortest geodesic
of $h_{ij}$, with length
$\ell_h=\min_\gamma\mathcal L_h$.
Together, these conditions give the optimal protocol,
minimum work, and its ratio to a stationary reference
$a(t)=a_0$:
\begin{equation}
\dot s_h=\frac{\ell_h}{a^2(t)I_\tau},
\qquad
\mathcal W_\lambda^{\min}=\frac{\ell_h^2}{I_\tau},
\qquad
\frac{\mathcal W_\lambda^{\min}}{\mathcal W_0^{\min}}
=\frac{\tau}{a_0^2I_\tau}.
\label{eq:opt}
\end{equation}
Here $s_h(t)$ is the accumulated arc length along
the optimal path, with $s_h(0)=0$ and
$s_h(\tau)=\ell_h$, and
$\mathcal W_0^{\min}=a_0^2\ell_h^2/\tau$
is the minimum work for the stationary reference.
Hence, the reference metric fixes the optimal spatial
path, while the background history determines how
that path is traversed and how much dissipation it
incurs. The minimum work is reduced,
$\mathcal W_\lambda^{\min}<\mathcal W_0^{\min}$,
whenever $a_0^2I_\tau>\tau$.
In particular, any nontrivial cyclic modulation with
$a(0)=a(\tau)=a_0$ and $a(t)\leq a_0$ throughout
the protocol lowers the minimum driving cost.
More generally, the minimum dissipation is governed
by the harmonic mean of the metric prefactor $a^2(t)$
over the entire background history, rather than by
its endpoint values; cyclicity alone does not ensure
a reduction.

If $\alpha$ responds dynamically to $\bm\lambda$,
its history becomes protocol dependent and must be
included as a dynamical constraint in the optimization
\emph{(Appendix~A)}. Perturbative
corrections for weak background modulation, including
nonseparable metrics, are derived in \emph{Appendix~B}.

\emph{Thermodynamic Hubble relation and
anomalous relaxation.--}
Having established how the temporal connection
modifies optimal driving, we now examine its
role in spontaneous relaxation, where external
controls remain fixed while the background
evolves [Fig.~\ref{fig:geometry_evolution}
(lower)].
Consider an observable $y=y[p]$ of the
probability distribution $p(z,t)$, whose
reduced dynamics admits an overdamped
gradient-flow description
$\dot y=-M(y,\alpha)\partial_y
\mathcal F(y;\lambda)$,
where $M>0$ is the mobility and
$\mathcal F$ is a free-energy landscape
independent of $\alpha$
~\cite{Onsager1931,Doi2011}.
The dissipation identity
$-\dot{\mathcal F}=G(y,\alpha)\dot y^2$
defines the metric $G=M^{-1}$.
When slow driving shifts the equilibrium
observable $y_{\rm eq}(\lambda)$,
the control-space friction metric follows
from the pullback
$g_{\lambda\lambda}
=G(y_{\rm eq},\alpha)
(\partial_\lambda y_{\rm eq})^2$.
Thus, within this reduced description,
driven control and spontaneous relaxation
share the same underlying dissipative geometry.

To characterize geometric evolution during
relaxation, consider an observable decreasing
monotonically toward a fixed target $y_f$.
Its instantaneous thermodynamic distance is
$R(t)=\int_{y_f}^{y(t)}
\sqrt{G(q,\alpha(t))}\,dq$,
evaluated at the current background state.
Differentiation yields the thermodynamic
Hubble relation
\begin{equation}
\dot R=H_RR-v_g,\quad
H_R=\frac{\dot\alpha}{2R}
\int_{y_f}^{y}dq\,
\frac{\pd_\alpha G(q,\alpha)}{\sqrt{G(q,\alpha)}},
\label{eq:hubble}
\end{equation}
where $v_g=-\sqrt{G(y,\alpha)}\dot y>0$
is the geometric relaxation speed, while
$H_RR$ describes the contribution from
metric evolution.
Here $H_R$ is the distance-weighted average
of the local temporal connection
$\frac12\pd_t\ln G$.
For a separable metric $G=a^2(t)h(y)$,
one obtains $H_R=H=\dot a/a$.
The same expansion rate that enters the
optimal-control geodesic
Eq.~(\ref{eq:geodesicH}) thus also governs
thermodynamic distance during spontaneous
relaxation.

As illustrated by the shrinking slide in Fig.~\ref{fig:geometry_evolution} (lower), the system spontaneously relaxes toward equilibrium with its environment, while geometric contraction further shortens the remaining thermodynamic distance. Conversely, geometric expansion stretches this distance and can counteract its decrease during relaxation. For a stationary metric, $H_R=0$ and $\dot R=-v_g<0$, whereas sufficiently strong
expansion ($H_RR>v_g$) can make $\dot R>0$
even when $\dot y<0$.
In the separable gradient-flow case,
expansion also reduces mobility, potentially
delaying or, if sufficiently rapid, preventing
arrival at a prescribed target.
An increasing thermodynamic distance,
however, need not imply an increase in
nonequilibrium free energy, nor does the
geometric expansion rate alone establish
anomalous relaxation.

The thermodynamic Hubble relation suggests that evolving geometry may influence relaxation differently for different initial preparations. This naturally raises the possibility of anomalous relaxation-time ordering, as in the Mpemba effect, where a system initially farther from equilibrium reaches a prescribed target earlier than one initially closer~\cite{TezaEtAl2026,LuRaz2017,Klich2019,
Megias2022,LinTuMa2026}. For monotonically relaxing observables,
this requires a reversal of their initial
ordering.
To quantify this effect, consider autonomous
dynamics $\dot x^A=F^A(x)$, with
$x^A=(y,\alpha)$, and initial states
$x(0;\chi)$ labeled so that larger $\chi$
corresponds to a greater initial displacement
from the target.
Differentiating the arrival condition
$y(\tau_f;\chi)=y_f$ yields
$d\tau_f/d\chi=-J(\tau_f)/\dot y(\tau_f)$,
where $J(t)=\partial_\chi y(t;\chi)$.
For monotonic relaxation with $\dot y<0$,
$J(\tau_f)<0$ implies $d\tau_f/d\chi<0$,
so a more distant initial state reaches
the target earlier.
Such reversals are excluded in scalar
dynamics with a common prescribed background,
but can arise when the background retains
information about initial preparation or
dynamically responds to the observable.
The initial-state dependence of relaxation
times is analyzed further in
\emph{Appendix~C}.

\emph{Application I: Prescribed friction and optimal control.--}
Consider an overdamped Brownian particle in a
harmonic trap $E(q;k)=kq^2/2$, with stiffness
$k(t)>0$ as the accessible control and environmental
friction $\Gamma(\alpha)$ as a hidden kinetic
parameter. At fixed bath temperature $T_b$,
its dynamics is
$\Gamma(\alpha)\dot q=-kq
+\sqrt{2\kb T_b\Gamma(\alpha)}\xi(t)$,
where $\xi(t)$ is Gaussian white noise
~\cite{SchmiedlSeifert2007,Zulkowski2012,
ZulkowskiOverdamped2015}.
For frozen parameters, the equilibrium distribution
is Gaussian with variance $\kb T_b/k$.
Applying Eq.~(\ref{eq:metric}) to $X_k=-q^2/2$
gives $g_{kk}(k,\alpha)
=\Gamma(\alpha)\kb T_b/(4k^3)$.
The equilibrium Fisher information
$\mathcal I_{kk}=1/(2k^2)$ is independent of
$\alpha$, whereas the force-correlation time is
$\Gamma/(2k)$. Thus, environmental evolution
changes the dissipative geometry without altering
the frozen equilibrium statistics.

For an arbitrary $\alpha(t)$,
the friction $\Gamma(t)=\Gamma[\alpha(t)]$
gives $H=\dot\Gamma/(2\Gamma)$.
The geodesic equation~(\ref{eq:geodesicH})
then becomes
\begin{equation}
\ddot k-\frac{3\dot k^2}{2k}
+\frac{\dot\Gamma}{\Gamma}\dot k=0.
\label{eq:kgeodesic}
\end{equation}
With $z=k^{-1/2}$, this equation reduces
to $d(\Gamma\dot z)/dt=0$.
Integration with fixed endpoints gives the optimal
protocol within the quadratic dissipation theory,
\begin{equation}
k_*(t)=
\left[
k_i^{-1/2}
+\left(k_f^{-1/2}-k_i^{-1/2}\right)
\mathcal S(t)
\right]^{-2},
\label{eq:koptimal}
\end{equation}
where $\mathcal S(t)\equiv
[\int_0^t ds\,\Gamma^{-1}(s)]/
[\int_0^\tau ds\,\Gamma^{-1}(s)]$.
For constant friction, $\mathcal S(t)=t/\tau$,
recovering the stationary geodesic protocol
with $k^{-1/2}$ linear in time.
Geodesic optimization of harmonic Brownian systems
also underlies shortcuts to isothermality, in which
auxiliary control maintains instantaneous equilibrium
~\cite{LiChenSunDong2022}.
Here, without such auxiliary control, evolving
friction instead changes the temporal allocation
of the optimal motion toward intervals of lower
resistance.

\begin{figure}[tbh!]
\centering
\includegraphics[width=\columnwidth]
{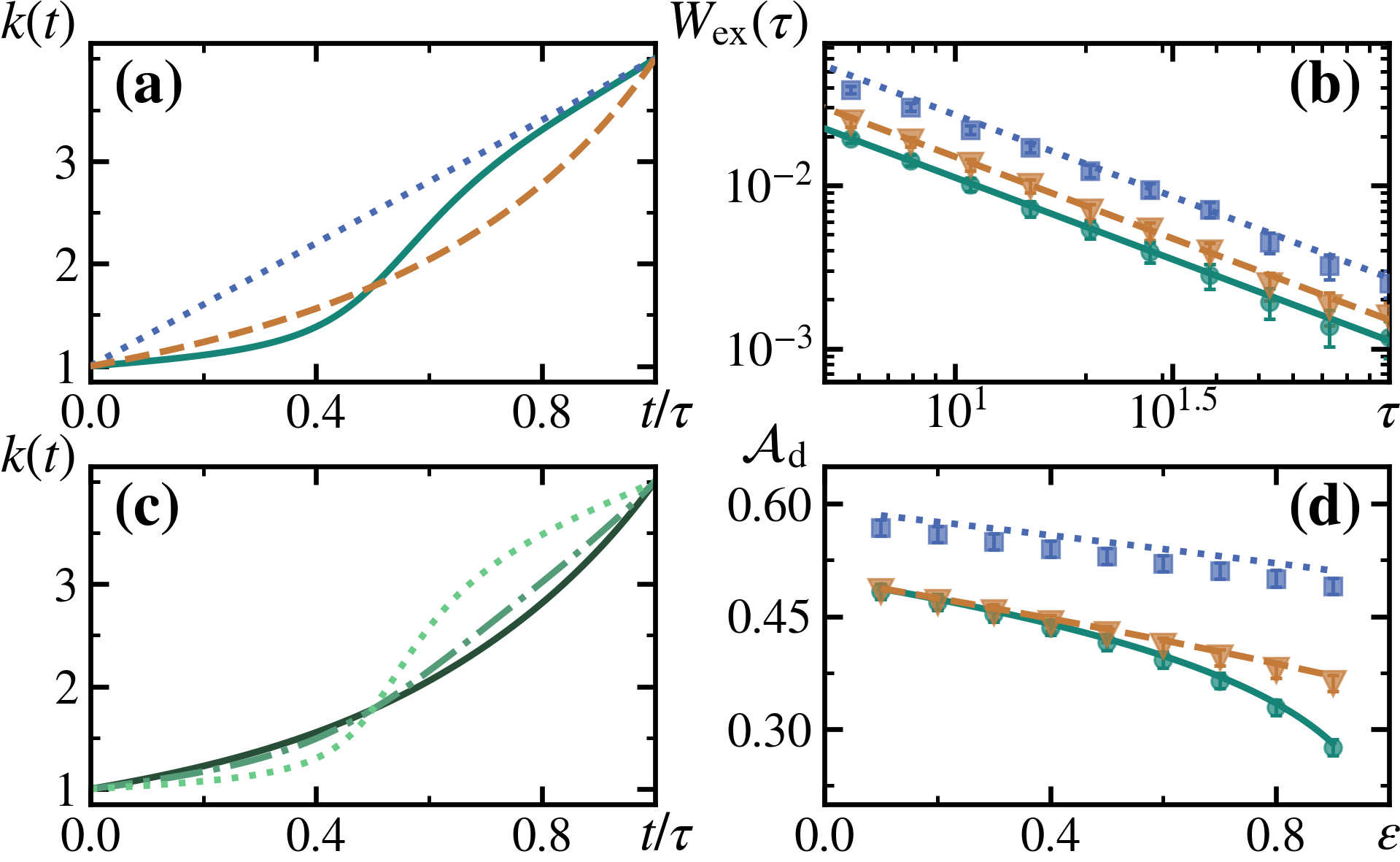}
\caption{Optimal control and dissipation under
periodically modulated friction.
(a) Stiffness protocols:
linear (blue dotted),
stationary-metric optimal (orange dashed),
and nonstationary optimal (green solid).
(b) Corresponding excess work versus
duration $\tau$, with $\varepsilon=0.8$
in (a,b).
(c) Nonstationary optimal protocols for
$\varepsilon=0.1$ (dark green solid),
$0.5$ (green dash-dotted), and
$0.9$ (light green dotted).
(d) Dissipation amplitudes
$\mathcal A_{\rm d}^{p}$ versus $\varepsilon$.
Simulation values are estimated by
$\widehat{\mathcal A}_{\rm d}^{p}
=[N_\tau^{-1}\sum_{j=1}^{N_\tau}
\tau_j\mathcal W_p(\tau_j)]^{1/2}$
over discrete durations
$10\leq\tau_j\leq100$. Symbols denote Brownian simulations
averaged over $10^5$ trajectories,
and curves show analytical predictions, with $\Gamma_0=1$, $k_i=1$, $k_f=4$, and $k_BT_b=1$.}
\label{fig:periodic_friction}
\end{figure}

As an example, consider periodically
modulated friction
$\Gamma(t)=\Gamma_0[1-\varepsilon
\sin^2(\pi t/\tau)]$,
with $0<\varepsilon<1$, completing one
cycle over the protocol duration $\tau$~\footnote{
More generally, the modulation
$\sin^2(\pi\omega t/\tau)$ allows the
period to differ from $\tau$, so the
endpoint frictions need not coincide.
Within quadratic response, the ratio
$\mathcal W_k^{\min}/\mathcal W_0^{\min}$
approaches unity as $\omega\to0$
and $\sqrt{1-\varepsilon}$ as
$\omega\to\infty$, reflecting its
dependence on the full friction history
through $\int_0^\tau dt/\Gamma(t)$.
The general solution is given in
\emph{Appendix~D}.
The high-frequency limit is formal,
as the driving and background evolution
must remain slow relative to the local
relaxation rate $k/\Gamma$.}. The optimal time parameterization is
$\mathcal S(\tilde t)=\pi^{-1}
\operatorname{atan2}
[\sqrt{1-\varepsilon}\sin(\pi\tilde t),
\cos(\pi\tilde t)]$,
where $\tilde t=t/\tau$ and the phase
is continuously unwrapped from $0$ to $\pi$.
Figure~\ref{fig:periodic_friction}(a)
compares three stiffness protocols
connecting the same endpoints:
linear driving (blue dotted),
the stationary-metric optimum
(orange dashed), and the nonstationary
optimum (green solid).
The stationary-metric optimum corresponds
to $\mathcal S(\tilde t)=\tilde t$,
whereas the nonstationary optimum
shifts driving toward periods of
lower friction. From Eq.~(\ref{eq:opt}), the minimum
excess work relative to the
constant-friction reference satisfies
\begin{equation}
\frac{\mathcal W_k^{\min}}{\mathcal W_0^{\min}}
=\sqrt{1-\varepsilon},
\mathcal W_0^{\min}
=\frac{\Gamma_0\kb T_b
(k_f^{-1/2}-k_i^{-1/2})^2}{\tau}.
\label{eq:brownian}
\end{equation}
Thus, despite identical initial and final
frictions, metric evolution reduces
the minimum dissipation while retaining
the familiar $1/\tau$ scaling
~\cite{SalamonBerry1983,Crooks2007,Ma2020PRL}.
Figure~\ref{fig:periodic_friction}(b)
compares Brownian simulations (symbols)
with the slow-driving predictions (lines)
derived in \emph{Appendix~D}.
Under the same friction modulation,
the stationary-metric optimum
(orange triangles) dissipates
$\mathcal W_{\rm stat}
=(1-\varepsilon/2)\mathcal W_0^{\min}
>\mathcal W_k^{\min}$.
The nonstationary optimum (green circles)
achieves the lowest dissipation,
while linear driving (blue squares)
incurs the highest.

Figures~\ref{fig:periodic_friction}(c,d)
show how the strength of geometric
evolution, characterized by $\varepsilon$,
affects optimal driving and dissipation.
As $\varepsilon$ increases, the
nonstationary optimal protocol becomes
increasingly nonuniform
[Fig.~\ref{fig:periodic_friction}(c)].
To quantify the dissipation reduction,
we define the dissipation amplitude
$\mathcal A_{\rm d}^{p}
=\lim_{\tau\to\infty}
\sqrt{\tau\mathcal W_p}$
for each protocol $p$, with
$\mathcal W_{\rm opt}=\mathcal W_k^{\min}$.
The nonstationary and stationary-metric
optima satisfy
$\mathcal A_{\rm d}^{\rm opt}/
\mathcal A_{\rm d}^{\rm stat}
=1-\varepsilon^2/16+O(\varepsilon^3)$~\footnote{From \emph{Appendix~D}, the exact dissipation amplitudes are
$\mathcal A_{\rm d}^{\rm opt}/\mathcal A_0
=(1-\varepsilon)^{1/4}$
and
$\mathcal A_{\rm d}^{\rm stat}/\mathcal A_0
=\sqrt{1-\varepsilon/2}$,
where $\mathcal A_0
=\sqrt{\tau\mathcal W_0^{\min}}$
is the constant-friction reference amplitude.
}. Thus, although the optimal driving
schedule changes at first order,
the additional dissipation reduction
emerges only at second order,
explaining the near coincidence of
the two amplitudes at small
$\varepsilon$ in
Fig.~\ref{fig:periodic_friction}(d).
Unlike dissipation reduction through
shorter geodesics enabled by
higher-dimensional control in stationary
thermodynamic geometry
~\cite{ChenZhaiSunDong2023},
the reduction here arises from
redistributing driving in time
along the same spatial path.

\emph{Application II: Mpemba relaxation in evolving dissipative geometry.--}
We now hold the trap stiffness fixed at $k=k_0$
and allow the surrounding medium to evolve through
heat exchange. Following the phase-change reservoir
mechanism of Ref.~\cite{LinTuMa2026}, consider a
finite bath containing two coexisting phases at
temperature $T_b$. Heat released by the tracer
converts a low-mobility phase into a high-mobility
phase, while latent heat maintains the bath
temperature. The hidden variable $\alpha(t)$ denotes
the fraction of the high-mobility phase; we assume
it is self-averaging and initially identical across
preparations. Let the tracer initially be equilibrated
at $T_i>T_b$, and define its configurational-temperature
ratio $\theta=k_0\langle q^2\rangle/(\kb T_b)$,
with $\theta_i=T_i/T_b$.
Since the mean potential energy is $\kb T_b\theta/2$,
energy conservation gives
\begin{equation}
\Theta[\alpha(t)-\alpha_0]
=Q(t)=\frac{\kb T_b}{2}
[\theta_i-\theta(t)],
\label{eq:medium}
\end{equation}
where $\Theta$ is the latent heat for complete
phase conversion. With the linear mobility rule
$\mu(\alpha)=\mu_l(1-\alpha)+\mu_h\alpha$,
where $\mu_h>\mu_l>0$, this becomes
\begin{equation}
\frac{\mu(t)}{\mu_0}
=1+\kappa[\theta_i-\theta(t)],\quad
\kappa=\frac{(\mu_h-\mu_l)\kb T_b}
{2\Theta\mu_0}>0,
\label{eq:mobility}
\end{equation}
with $\mu_0=\mu(\alpha_0)=\Gamma_0^{-1}$.
All preparations therefore start with the same
mobility, but hotter ones can develop greater
mobility enhancement through their larger heat
release.

\begin{figure}[tbh!]
\centering
\includegraphics[width=\columnwidth]{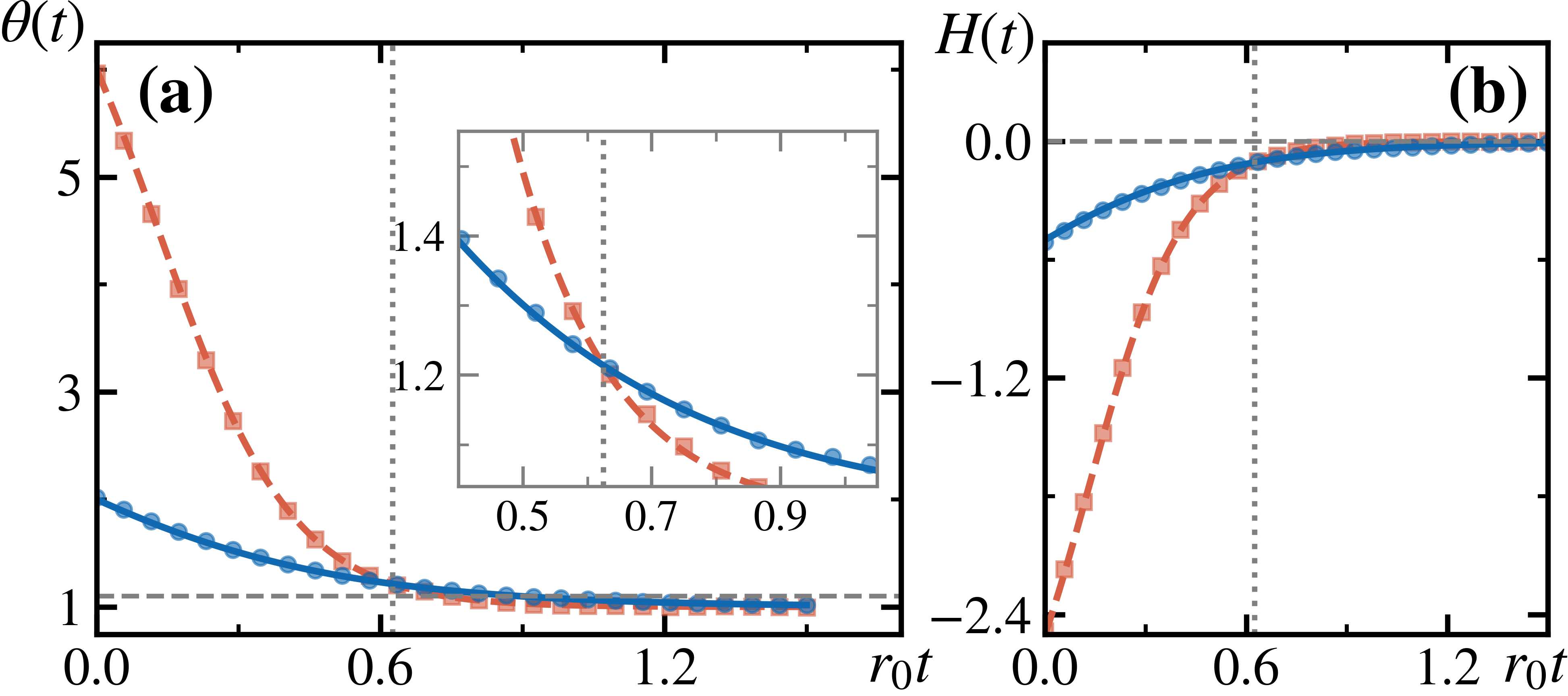}\caption{Mpemba relaxation with geometric contraction.
(a) Temperature relaxation from hot
($T_h=6T_b$, red) and cold
($T_c=2T_b$, blue) initial states.
The inset highlights the crossing
(vertical dotted line); the horizontal
dashed line marks $\theta_f=1.1$.
(b) Corresponding geometric expansion
rates $H_h(t)$ and $H_c(t)$.
Curves are analytical predictions,
and symbols are Brownian simulations
averaged over $10^5$ trajectories with $\kappa=0.5$, $r_0=1$.}

\label{fig:contracting_mpemba}
\end{figure}

In the self-averaging approximation, the tracer distribution remains Gaussian,
with variance $\sigma^2=\langle q^2\rangle$
and dissipative metric
$G_{\sigma\sigma}=\Gamma(t)=\mu^{-1}(t)$.
The corresponding pullback
$g_{kk}=\Gamma(t)\kb T_b/(4k^3)$
recovers the conditional friction metric
of Application I
\emph{(Appendix~D)}.
The configurational temperature
$\theta=k_0\sigma^2/(\kb T_b)$
obeys
$\dot\theta=-2r_0
[1+\kappa(\theta_i-\theta)](\theta-1)$,
where $r_0=k_0/\Gamma_0$.
Defining $A_i=1+\kappa(\theta_i-1)$,
we obtain the exact solution
\begin{equation}
\theta(t)=1+
\frac{A_i(\theta_i-1)}
{e^{2r_0A_it}+\kappa(\theta_i-1)}.
\label{eq:relaxsolution}
\end{equation}
Without mobility feedback ($\kappa=0$),
relaxation is exponential and preserves
the initial temperature ordering. For $\kappa>0$, preparation-dependent mobility can induce a Mpemba crossing. As shown in Fig.~\ref{fig:contracting_mpemba}(a), the initially hotter system
($T_h=6T_b$, red) cools below the colder one ($T_c=2T_b$, blue) at $r_0t_\times\simeq0.62$ (inset), reaching $\theta_f=1.1$ earlier ($r_0\tau_h\simeq0.74$ versus $r_0\tau_c\simeq0.89$). Brownian simulations (symbols) agree with the analytical solution
[Eq.~(\ref{eq:relaxsolution}), curves].

The evolving mobility also determines
the geometry of relaxation.
For a fixed target width
$\sigma_f<\sigma(t)$,
the thermodynamic distance
$R=\sqrt{\Gamma(t)}(\sigma-\sigma_f)$
obeys the Hubble relation
$\dot R=HR-v_g$, where
$v_g=-\sqrt{\Gamma(t)}\dot\sigma>0$.
Since $\Gamma=\mu^{-1}(\alpha)$,
the expansion rate is
$H(t)=-r_0\kappa[\theta(t)-1]<0$
\emph{(Appendix~D)}.
Thus, heat-induced phase conversion
increases mobility and contracts
the dissipative geometry.
As shown in Fig.~\ref{fig:contracting_mpemba}(b),
the hotter preparation undergoes
stronger geometric contraction
($H_h<H_c<0$) before the crossing,
accompanying its accelerated relaxation.
Importantly, geometric contraction
alone does not guarantee a Mpemba effect;
the reversal of relaxation ordering
depends on the full
preparation-dependent mobility dynamics.

\emph{Concluding remarks.--}We have established a framework of
nonstationary thermodynamic geometry
in which evolving internal or environmental
degrees of freedom reshape dissipative
response, even when equilibrium statistics
remain unchanged.
The temporal connection generalizes
minimum-dissipation geodesics and
governs the redistribution of optimal
excess power.
A thermodynamic Hubble relation further
describes how metric evolution affects
thermodynamic distance during spontaneous
relaxation.
Brownian examples demonstrate reduced
minimum dissipation and Mpemba-type
relaxation, with the latter requiring
preparation-dependent dynamics rather
than geometric contraction alone.

A recent study developed stochastic
thermodynamics on evolving physical
manifolds with time-dependent metrics
~\cite{Nagase2026}.
Although physical-space geometry is
distinct from thermodynamic dissipative
geometry, spatial deformation may alter
relaxation dynamics and force correlations,
thereby inducing an evolving friction metric.
Applying our framework to such systems,
while accounting for the energetic costs
of spatial deformation, offers an
interesting direction for future research.
More broadly, our results provide a
geometric perspective on finite-time
processes in evolving environments
~\cite{Ma2020FiniteReservoir,YuanMaSun2022,
Ma2023Polytropic,TarantoEtAl2025,
LipkaBartosikPerarnauLlobet2025}.
Another natural extension is the joint
optimization of external controls and
self-consistent background dynamics,
including their energetic costs.

\textit{Acknowledgment.---}This work is supported by
the National Natural Science Foundation under
Grant No. 12675036. The authors used ChatGPT (OpenAI)
for language polishing and assistance with coding
and debugging. The scientific question, conceptual
framework, and theoretical development were
independently formulated by the authors. The
authors take full responsibility for the Letter.

\end{CJK*}
\bibliography{references}
%\clearpage
\onecolumngrid{}
\begin{center}
    {\Large\bfseries End Matter}
\end{center}
\twocolumngrid{}

\emph{Appendix A: Conditional dissipation
and background dynamics.--}
The friction metric used in the main text
describes dissipation associated with the
accessible controls at a fixed instantaneous
background state. For the conjugate forces
$X_i=-\partial_{\lambda^i}E$, linear
response gives~\cite{SivakCrooks2012}
\begin{equation}
\delta\langle X_i(t)\rangle\simeq
-\beta\int_0^\infty ds\,
\langle\delta X_i(0)\delta X_j(s)
\rangle_{\bm\lambda,\alpha}
\dot\lambda^j(t-s).
\label{eq:responseapp}
\end{equation}
When the controls and background evolve
slowly compared with the force-correlation
times, the response can be approximated
locally, yielding the conditional
Green--Kubo metric~(\ref{eq:metric})
and excess power~(\ref{eq:power}).
This construction determines the
dissipation associated with
$\dot{\bm\lambda}$, but does not specify
the contributions from background motion
or its coupling to the controls.

To distinguish these contributions,
suppose a full friction tensor
$\mathbb G$ exists on the extended
coordinate space $(\bm\lambda,\alpha)$.
The quadratic dissipation then takes
the form
\begin{equation}
\mathcal W_{\rm diss}^{\rm tot}
=\int_0^\tau dt\,\big[\mathbb G_{ij}\dot\lambda^i\dot\lambda^j
+2\mathbb G_{i\alpha}\dot\lambda^i\dot\alpha+\mathbb G_{\alpha\alpha}\dot\alpha^2
\big].
\label{eq:fullcostapp}
\end{equation}
Holding the background fixed identifies
the conditional metric as
$g_{ij}=\mathbb G_{ij}$.
By contrast, minimizing the instantaneous
quadratic form over an unconstrained
$\dot\alpha$ gives
\begin{equation}
g_{ij}^{\rm rel}
=\mathbb G_{ij}
-\frac{\mathbb G_{i\alpha}\mathbb G_{\alpha j}}
{\mathbb G_{\alpha\alpha}},
\qquad \mathbb G_{\alpha\alpha}>0.
\label{eq:reduced_metric}
\end{equation}
This Schur complement generally differs
from the conditional metric and assumes
that the background velocity can be
adjusted freely. Neither this freedom
nor the additional components of
$\mathbb G$ follow from the conditional
Green--Kubo response alone.

For a prescribed background $\alpha(t)$,
the reduced optimization minimizes
$\mathcal W_\lambda$ in
Eq.~(\ref{eq:action}).
The Euler--Lagrange equation of
$L=g_{ij}(\bm\lambda,t)
\dot\lambda^i\dot\lambda^j$ reads
\begin{equation}
2g_{kj}\ddot\lambda^j+(2\partial_i g_{kj}-\partial_k g_{ij})
\dot\lambda^i\dot\lambda^j+2\partial_tg_{kj}\dot\lambda^j=0.
\label{eq:ELapp}
\end{equation}
Multiplying by $g^{\ell k}/2$ gives
the generalized geodesic
equation~(\ref{eq:geodesic}), while
contraction with the control velocity
yields the excess-power
relation~(\ref{eq:energy}).

If the background instead responds
dynamically to the controls through
$\dot\alpha=F(\alpha,\bm\lambda,t)$,
its trajectory is no longer prescribed
and must be included as a constraint.
Writing $u^i=\dot\lambda^i$,
the Pontryagin Hamiltonian for minimizing
the accessible driving cost is
\begin{equation}
\mathcal H
=g_{ij}(\bm\lambda,\alpha)u^iu^j
+p_i u^i+p_\alpha F(\alpha,\bm\lambda,t).
\label{eq:pmp}
\end{equation}
For a positive-definite metric and
unconstrained control velocities,
minimization gives
$u_*^i=-g^{ij}p_j/2$.
The costates obey
$\dot p_i
=-(\partial_i g_{jk})u^ju^k
-p_\alpha\partial_i F$
and
$\dot p_\alpha
=-(\partial_\alpha g_{ij})u^iu^j
-p_\alpha\partial_\alpha F$.
Together with the state equations and
boundary conditions, these relations
determine the optimal protocol while
accounting for the background response.
For a freely evolving terminal
background, the corresponding
transversality condition is
$p_\alpha(\tau)=0$.

The accessible driving cost need not
equal the total irreversible dissipation.
For example, if the background follows
an Onsager relation
$\dot\alpha=L_\alpha\mathcal A_\alpha$,
where $\mathcal A_\alpha$ is a
dimensionless thermodynamic affinity,
its dissipated power is
$\kb T_b\dot\alpha^2/L_\alpha$
~\cite{Onsager1931,Seifert2012}.
When cross-channel dissipation is
negligible, the total cost becomes
\begin{equation}
\mathcal W_{\rm diss}^{\rm tot}
=\int_0^\tau dt\left[
g_{ij}\dot\lambda^i\dot\lambda^j
+\frac{\kb T_b}{L_\alpha}\dot\alpha^2
\right].
\label{eq:totaldissapp}
\end{equation}
For the constrained dynamics
$\dot\alpha=F(\alpha,\bm\lambda,t)$,
minimizing this total cost amounts
to adding $\kb T_b F^2/L_\alpha$
to the running cost in
Eq.~(\ref{eq:pmp}).
An autonomously evolving background
remains subject to its dynamical
constraint, whereas an externally
controlled background may provide
an additional control variable,
provided its energetic cost is included.
When cross-channel responses are
significant, the full dissipation
tensor~(\ref{eq:fullcostapp}) must
be retained.

\emph{Appendix B: Perturbative corrections
to minimum dissipation.--}
We examine how weak prescribed background
modulation affects the minimum excess work,
including cases where the metric is not
conformally separable.
Let $\alpha(t)=\alpha_0+\epsilon\chi(t)$,
with $|\epsilon|\ll1$. Expanding around
$\alpha_0$ gives
\begin{equation}
g_{ij}(\bm\lambda,t)
=g_{ij}^{(0)}(\bm\lambda)
+\epsilon\chi(t)b_{ij}(\bm\lambda)
+O(\epsilon^2),
\label{eq:metricpert}
\end{equation}
where
$g_{ij}^{(0)}=g_{ij}(\bm\lambda,\alpha_0)$
and $b_{ij}=\partial_\alpha
g_{ij}|_{\alpha_0}$. Assuming the optimal protocol varies smoothly
with $\epsilon$, let $\bm\lambda_0(t)$ minimize the
dissipation for the stationary metric
$g_{ij}^{(0)}$, and write the perturbed
optimal protocol as
$\bm\lambda_\epsilon
=\bm\lambda_0+\epsilon\bm\eta+O(\epsilon^2)$,
with $\bm\eta(0)=\bm\eta(\tau)=0$.
Since $\bm\lambda_0$ extremizes the
unperturbed action with fixed endpoints,
its first variation vanishes.
The correction to the optimal protocol
therefore contributes only at second order,
and expanding Eq.~(\ref{eq:action}) yields
\begin{equation}
\mathcal W_\lambda^{\min}
=\mathcal W_0^{\min}
+\epsilon\int_0^\tau dt\,\chi(t)
b_{ij}(\bm\lambda_0)
\dot\lambda_0^i\dot\lambda_0^j+O(\epsilon^2),
\label{eq:actionvariationapp}
\end{equation}
where
$\mathcal W_0^{\min}
=\mathcal W_0[\bm\lambda_0]$
is the minimum work for the stationary
reference.
Thus, the leading correction is determined
entirely by the unperturbed optimal protocol.
For a nonseparable metric, the directional
response
$b_{ij}\dot\lambda_0^i\dot\lambda_0^j$
generally varies along the path, so even
a zero-mean modulation
$\langle\chi\rangle_t=0$ can change
the minimum work at first order, where
$\langle f\rangle_t
=\tau^{-1}\int_0^\tau f(t)\,dt$.

For the conformally separable metric
$g_{ij}=a^2(\alpha)h_{ij}(\bm\lambda)$,
the metric response simplifies to
$b_{ij}=c\,g_{ij}^{(0)}$. Expanding the metric scale factor gives
$a^2[\alpha(t)]/a_0^2
=1+\epsilon c\chi(t)
+\epsilon^2d\chi^2(t)+O(\epsilon^3)$,
where $a_0=a(\alpha_0)$,
$c=a_0^{-2}\partial_\alpha a^2|_{\alpha_0}$,
and $d=(2a_0^2)^{-1}
\partial_\alpha^2a^2|_{\alpha_0}$.
Substituting into Eq.~(\ref{eq:opt})
and expanding to second order yields
\begin{equation}
\frac{\mathcal W_\lambda^{\min}}
{\mathcal W_0^{\min}}
\simeq1+\epsilon c\langle\chi\rangle_t
+\epsilon^2\Big[
(d-c^2)\langle\chi^2\rangle_t
+c^2\langle\chi\rangle_t^2
\Big].
\label{eq:workpert}
\end{equation}
Unlike the general case, the first-order
correction depends only on the mean
background modulation.
For $\langle\chi\rangle_t=0$, the leading
correction is
$\epsilon^2(d-c^2)\langle\chi^2\rangle_t$,
so the minimum dissipation decreases
when $d<c^2$.
In particular, a locally linear metric
response ($d=0$, $c\ne0$) always lowers
the minimum work under nontrivial
zero-mean modulation, whereas nonlinear
dependence on $\alpha$ can reverse
the effect.

\emph{Appendix C: Initial-state dependence
of relaxation times.--}
We consider autonomous relaxation
$\dot x^A=F^A(x)$, where
$x^A=(y,\alpha)$ includes an observable
$y$ and an evolving background
variable $\alpha$.
For a target surface
$\Sigma_f=\{x:y=y_f\}$,
let $\tau_f(x)$ denote the first time
a trajectory starting from $x$ reaches
the target. Provided this time is finite
and the crossing is transverse,
the first-passage time satisfies
the backward transport equation $F^A\partial_A\tau_f=-1,\tau_f|_{\Sigma_f}=0.$ This equation provides a global
description of relaxation times
over the space of initial states.
To examine how the arrival time depends
on preparation, consider a family of
initial conditions
$(y_i(\chi),\alpha_i(\chi))$,
where increasing $\chi$ corresponds
to a larger initial displacement of $y$
from the target.
Writing the relaxation dynamics as
$\dot y=f(y,\alpha)$ and
$\dot\alpha=F(y,\alpha)$,
the sensitivities
$J(t)=\partial_\chi y(t;\chi)$
and $K(t)=\partial_\chi\alpha(t;\chi)$
satisfy
\begin{equation}
\dot J=f_yJ+f_\alpha K,\qquad
\dot K=F_yJ+F_\alpha K.
\label{eq:genericvar}
\end{equation}
Their initial values are
$J(0)=y_i'(\chi)$ and
$K(0)=\alpha_i'(\chi)$.
Differentiating the arrival condition
$y(\tau_f;\chi)=y_f$ gives
\begin{equation}
\frac{d\tau_f}{d\chi}
=-\frac{J(\tau_f)}
{f[y_f,\alpha(\tau_f)]},
\label{eq:arrivalsensitivityapp}
\end{equation}
which relates the initial-state dependence
of arrival times to the observable sensitivity.
For monotonic cooling with $f<0$,
a negative $J(\tau_f)$ indicates
that a more distant initial state
reaches the target earlier.

These equations also clarify the role
of background memory.
If all preparations share the same
initial background state, then
$K(0)=0$.
When the background evolves independently
of $y$, so that $F_y=0$, one has
$K(t)=0$ and
$\dot J=f_yJ$, which preserves the
sign of $J$.
Likewise, if $f_\alpha=0$, the
observable evolves independently
of the background and its ordering
cannot reverse.
Thus, with identical initial backgrounds,
a reversal requires coupling in both
directions between $y$ and $\alpha$, 
which is illustrated by Application II.
If the initial background instead
depends on preparation,
$K(0)=\alpha_i'(\chi)\ne0$,
the term $f_\alpha K$ can reverse
the observable ordering even when
$F_y=0$.
In this case, the preparation-dependent
background retains information about
the initial state and modifies the
subsequent relaxation.

\emph{Appendix D: Brownian dynamics and dissipative geometry.--}
We provide the calculations underlying the two
Brownian applications and establish their common
dissipative geometry. For Application I, the conjugate force is $X_k=-q^2/2$.
At fixed $(k,\alpha)$, the equilibrium correlation
$\langle q(0)q(s)\rangle=(\kb T_b/k)e^{-ks/\Gamma}$
and Wick's theorem give $\langle\delta X_k(0)\delta X_k(s)\rangle
=(\kb T_b)^2e^{-2ks/\Gamma}/(2k^2).$ Substitution into the conditional Green--Kubo
formula~(\ref{eq:metric}) yields
\begin{equation}
g_{kk}=\frac{\kb T_b}{2k^2}
\int_0^\infty e^{-2ks/\Gamma}ds
=\frac{\Gamma\kb T_b}{4k^3}.
\label{eq:brownmetricapp}
\end{equation}
Thus, friction enters the metric through
the correlation time, without changing
the equilibrium fluctuations.

For an arbitrary prescribed history
$\Gamma(t)=\Gamma[\alpha(t)]$, the spatial
and temporal connections are
$\Gamma^k_{kk}=\frac12\partial_k\ln g_{kk}
=-3/(2k)$ and
$2\mathsf H^k{}_k=\partial_t\ln g_{kk}
=\dot\Gamma/\Gamma$, respectively,
with derivatives taken at fixed $t$ and $k$.
The resulting equation is precisely the
conformally separable geodesic
Eq.~(\ref{eq:geodesicH}), and reproduces
Eq.~(\ref{eq:kgeodesic}).
Introducing $z=k^{-1/2}$ eliminates the
spatial connection and reduces the equation
to $d(\Gamma\dot z)/dt=0$.
With fixed endpoints and
$J_\tau=\int_0^\tau dt/\Gamma(t)$,
integration recovers the optimal stiffness
protocol~(\ref{eq:koptimal}). For the
normalized metric scale $a^2(t)=\Gamma(t)/\Gamma_0$,
the main-text quantity satisfies
$I_\tau=\Gamma_0 J_\tau$.

To obtain the analytical reference curves
in Fig.~\ref{fig:periodic_friction}(b),
we consider two protocols under the same
friction modulation
$\Gamma(t)=\Gamma_0[1-\varepsilon
\sin^2(\pi t/\tau)]$.
Using $\mathcal W_k=\kb T_b
\int_0^\tau dt\,\Gamma(t)\dot z^2$,
the linear protocol
$k_{\rm lin}(t)=k_i+\Delta k\,t/\tau$
and the stationary-metric optimum
$z_{\rm stat}(t)=z_i+\Delta z\,t/\tau$
yield
\begin{equation}
\begin{aligned}
\frac{\mathcal W_{\rm lin}}{\mathcal W_0^{\min}}=\frac{(\Delta k)^2}{4(\Delta z)^2}
\int_0^1du\,
\frac{1-\varepsilon\sin^2(\pi u)}
{(k_i+\Delta k\,u)^3},\frac{\mathcal W_{\rm stat}}{\mathcal W_0^{\min}}
&=1-\frac{\varepsilon}{2},
\end{aligned}
\label{eq:referenceworkapp}
\end{equation}
where $\Delta k=k_f-k_i$,
$\Delta z=k_f^{-1/2}-k_i^{-1/2}$,
and $\mathcal W_0^{\min}
=\Gamma_0\kb T_b(\Delta z)^2/\tau$.
These results confirm that
$\mathcal W_k^{\min}<\mathcal W_{\rm stat}
\leq\mathcal W_{\rm lin}$ for $0<\varepsilon<1$, demonstrating the advantage of accounting for metric evolution in optimal control.

We now consider a periodic friction modulation
with an arbitrary frequency,
\begin{equation}
\Gamma(t)=\Gamma_0
[1-\varepsilon\sin^2(\pi\omega\tilde t)],
\qquad \tilde t=t/\tau,
\label{eq:generalcycleapp}
\end{equation}
where $0<\varepsilon<1$ and $\omega>0$.
Its period $\tau/\omega$ need not equal
the operation time $\tau$, giving a final
friction $\Gamma_f=\Gamma_0
[1-\varepsilon\sin^2(\pi\omega)]$
that generally differs from $\Gamma_0$.
To evaluate the optimal time parameterization,
define $r=\sqrt{1-\varepsilon}$ and
$\varphi_\varepsilon(u)=
\operatorname{atan2}(r\sin u,\cos u)$,
with the phase continuously unwrapped so that
$\varphi_\varepsilon(0)=0$ and
$\varphi_\varepsilon(u+\pi)=
\varphi_\varepsilon(u)+\pi$.
Since $d\varphi_\varepsilon/du
=r/(1-\varepsilon\sin^2u)$, integration gives
\begin{equation}
\begin{aligned}
J_\tau&=\frac{\tau\varphi_\varepsilon(\pi\omega)}
{\Gamma_0\pi\omega r},\qquad
\mathcal S(\tilde t)
=\frac{\varphi_\varepsilon(\pi\omega\tilde t)}
{\varphi_\varepsilon(\pi\omega)}.
\end{aligned}
\label{eq:cycleintegralapp}
\end{equation}
Substituting $\mathcal S(\tilde t)$ into
Eq.~(\ref{eq:koptimal}) gives the optimal
stiffness control $k_*(t)$ for arbitrary
$\omega$. With $z_*(t)=k_*^{-1/2}(t)$ and
$\dot z_*(t)=\Delta z/[\Gamma(t)J_\tau]$,
the excess power and minimum work follow
from $\mathcal W_k=\kb T_b
\int_0^\tau\Gamma(t)\dot z^2dt$:

\begin{equation}
P_{\rm ex}^{(k)}(t)=\frac{\kb T_b(\Delta z)^2}{\Gamma(t)J_\tau^2},
\quad
\eta\equiv\frac{\mathcal W_k^{\min}}{\mathcal W_k^{\min}[\Gamma_0]}=
\frac{\pi\omega r}{\varphi_\varepsilon(\pi\omega)},
\label{eq:brownpower}
\end{equation}
where $\mathcal W_k^{\min}[\Gamma_0]
=\Gamma_0\kb T_b(\Delta z)^2/\tau$
is the minimum excess work for constant
friction $\Gamma_0$.
The optimal power satisfies
$\dot P_{\rm ex}^{(k)}
=-(\dot\Gamma/\Gamma)P_{\rm ex}^{(k)}$,
consistent with Eq.~(\ref{eq:energy}). The frequency-dependent factor in
Eq.~(\ref{eq:brownpower}) reflects the entire
friction history rather than its final
value $\Gamma_f$ alone.
For $\sin(\pi\omega)\ne0$, the result can
also be parameterized by $\Gamma_f$ through
$\varepsilon=(1-\Gamma_f/\Gamma_0)/
\sin^2(\pi\omega)$.
For integer or half-integer $\omega>0$,
the unwrapped phase satisfies
$\varphi_\varepsilon(\pi\omega)=\pi\omega$,
giving $\eta=\sqrt{1-\varepsilon}$.
The minimum work is therefore identical
despite different final frictions:
$\Gamma_f=\Gamma_0$ for integer $\omega$,
but $\Gamma_f=\Gamma_0(1-\varepsilon)$
for half-integer $\omega$.
In particular, $\omega=1$ recovers the
optimal time parameterization and minimum work~(\ref{eq:brownian}) in Application I.

For weak modulation, the optimal time
parameterization becomes
$\mathcal S(\tilde t)=\tilde t+
\varepsilon[\tilde t\sin(2\pi\omega)
-\sin(2\pi\omega\tilde t)]/(4\pi\omega)
+O(\varepsilon^2)$, while
\begin{equation}
\eta=1-\varepsilon\left[
\frac12-\frac{\sin(2\pi\omega)}{4\pi\omega}
\right]+O(\varepsilon^2).
\label{eq:clockpertapp}
\end{equation}
For fixed $\varepsilon$, the exact result
further gives the frequency limits
\begin{equation}
\eta
=
\begin{cases}
1-\dfrac{\varepsilon\pi^2\omega^2}{3}
+O(\omega^4), & \omega\to0,\\[4pt]
\sqrt{1-\varepsilon}+O(\omega^{-1}),
& \omega\to\infty.
\end{cases}
\label{eq:frequencylimitsapp}
\end{equation}
At low frequency, the nearly constant
friction recovers the stationary minimum
work. At high frequency, the integrated
inverse friction approaches its period
average, while $\mathcal S(\tilde t)
=\tilde t+O(\omega^{-1})$.
The latter limit requires sufficiently
slow dynamics: both the modulation rate
$\omega/\tau$ and the relative stiffness
variation rate must remain small compared
with the relaxation rate $k/\Gamma$.

For Application II, in the self-averaging
approximation, the overdamped Brownian tracer
obeys $dq=-k_0\mu(\alpha)q\,dt
+\sqrt{2\kb T_b\mu(\alpha)}\,dW_t$,
where $\alpha(t)$ denotes the fraction of
the high-mobility phase in the finite reservoir.
For a centered Gaussian initial state, the
distribution remains Gaussian, with variance
$v=\langle q^2\rangle$ satisfying
$\dot v=-2\mu(\alpha)(k_0v-\kb T_b)$.
Since the trap stiffness remains fixed,
no external work is performed, and energy
conservation gives the cumulative heat
transferred to the reservoir as
$Q(t)=k_0[v(0)-v(t)]/2$.
Equating this heat to the latent-heat change
$\Theta[\alpha(t)-\alpha_0]$
gives Eq.~(\ref{eq:medium}), while the
linear mixture rule
$\mu(\alpha)=\mu_l(1-\alpha)+\mu_h\alpha$
yields Eq.~(\ref{eq:mobility}).

The coupled dynamics admits an exact solution.
Defining $x=\theta-1$, $x_i=\theta_i-1$,
$r_0=k_0/\Gamma_0$, and
$A_i=1+\kappa x_i$,
Eq.~(\ref{eq:mobility}) gives
$\mu(\alpha)/\mu_0=A_i-\kappa x$.
The variance dynamics thus reduces to
$\dot x=-2r_0(A_i-\kappa x)x$.
Direct integration yields
\begin{equation}
\frac{x(t)}{A_i-\kappa x(t)}
=x_i e^{-2r_0A_it},
\end{equation}
which gives the exact relaxation solution
(\ref{eq:relaxsolution}). To identify the dissipative geometry
underlying this relaxation, we write
$v=\sigma^2$ and
$\sigma_b^2=\kb T_b/k_0$.
For the Gaussian tracer, the mean potential
energy is $U=k_0\sigma^2/2$, while the
configurational entropy is
$S=\kb\ln\sigma+\mathrm{const}$.
The nonequilibrium free energy
$\mathcal F=U-T_bS$ is therefore, up to
an additive constant
~\cite{Seifert2012,EspositoVdB2011},
$\mathcal F(\sigma)=k_0\sigma^2/2
-\kb T_b\ln(\sigma/\sigma_b)$.
The variance dynamics can be written
as the gradient flow
$\dot\sigma=-\Gamma^{-1}
\partial_\sigma\mathcal F$,
where $\Gamma=\mu^{-1}(\alpha)$.
Consequently,
$-\dot{\mathcal F}=\Gamma\dot\sigma^2$,
identifying the dissipative metric
$G_{\sigma\sigma}=\Gamma(\alpha)$.
For slowly driven stiffness $k$, its
pullback along
$\sigma_{\rm eq}(k)=\sqrt{\kb T_b/k}$
gives $g_{kk}=\Gamma\kb T_b/(4k^3)$,
consistent with the frozen Green--Kubo
result~(\ref{eq:brownmetricapp}).
For spontaneous cooling ($\sigma>\sigma_b$),
the geometric distance is
$R=\sqrt{\Gamma}(\sigma-\sigma_b)$,
which directly satisfies the thermodynamic
Hubble relation~(\ref{eq:hubble}) with
$H=\frac12 d\ln\Gamma/dt$.
Using the reservoir dynamics gives
\begin{equation}
H(t)=\frac12\frac{d\ln\Gamma}{dt}
=-\frac{\mu_h-\mu_l}{2\mu(\alpha)}
\dot\alpha
=-r_0\kappa[\theta(t)-1].
\label{eq:contractionrate}
\end{equation}
For $\theta>1$ and $\kappa>0$,
heat absorption increases the reservoir
mobility, resulting in geometric
contraction ($H<0$).

Finally, the constant-temperature
reservoir description is valid only
during two-phase coexistence,
$0<\alpha(t)<1$.
For the parameters in
Fig.~\ref{fig:contracting_mpemba},
$\kappa=0.5$ can be realized with
$\alpha_0=0.05$, $\mu_h/\mu_l=21$,
and $\Theta=10\kb T_b$ per tracer.
The final phase fractions are
$\alpha_h=0.30$ and $\alpha_c=0.10$,
both within the coexistence region.
The self-averaging approximation neglects
fluctuations of the reservoir composition,
which would require a joint stochastic
description.

\end{document}